\documentclass[12pt]{article}

\usepackage{epsf}
\usepackage[dvips]{graphicx}

\newcommand{\vecc}[1]{\mbox{\boldmath $#1$}}
\def\ex{\hbox{e}}

\def\e{\epsilon}

\def\pd{\partial}

\def\z{\zeta}

\def\ex{\hbox{e}}
\def\S{\Sigma}

\def\<{\langle}
\def\>{\rangle}

\def\a{\alpha}

\def\d{\delta}

\def\m{\mu}
\def\n{\nu}

\def\z{\zeta}

\def\({\left(}
\def\[{\left[}
\def\){\right)}
\def\]{\right]}

\def\pd{\partial}

\def\pa{{\cal P}}

\def\halb{\frac{1}{2}}

\begin{document}
\hfill{RUB-TPII-12/08}
\bigskip

\begin{center}
{\bfseries RECENT PROGRESS \\ IN TRANSVERSE-MOMENTUM DEPENDENT
           PDFs\footnote{Invited talk presented by the first author at
           XIX International Baldin Seminar on High Energy Physics
           Problems ``Relativistic Nuclear Physics and Quantum
           Chromodynamics'', Dubna, Russia, September 29 to October 4,
           2008.}}

\vskip 5mm

I.O. Cherednikov$^{1,2 *}$ and N.G. Stefanis$^{3 \dag}$

\vskip 5mm

{\small
(1) {\it
Bogoliubov Laboratory of Theoretical Physics, JINR,
RU-141980 Dubna, Russia
}
\\
(2) {\it
INFN Gruppo collegato di Cosenza, I-87036 Rende, Italy
}
\\
(3) {\it
Institut f\"{u}r Theoretische Physik II,
Ruhr-Universit\"{a}t Bochum, D-44780 Bochum, Germany
}
\\
$^*$ {\it
E-mail: igor.cherednikov@jinr.ru
}
\\
$^\dag$ {\it Email: stefanis@tp2.ruhr-uni-bochum.de }
}
\end{center}

\vskip 5mm

\begin{center}
\begin{minipage}{150mm}
\centerline{\bf Abstract}
We present some recently obtained results on transverse-momentum
dependent parton distribution functions (TMD PDFs), the latter being
important ingredients for the description of semi-inclusive hadronic
processes.
Special attention is payed to the renormalization group (RG) properties
of these objects.
In particular, their leading-order anomalous dimension is calculated in
the light-cone gauge.
It is shown that the RG properties can be used to reveal, in the most
economic way, the general structure of the gauge links entering the
operator definition of TMD PDFs.
\end{minipage}
\end{center}

\vskip 10mm

\section{Introduction}
\label{sec:intro}

Path-ordered gauge links (Wilson lines) of the form
\begin{equation}
  { [y,x|\Gamma] }
=
  {\cal P} \exp
  \left[-ig\int_{x[\Gamma]}^{y}dz_{\mu} A_{a}^{\mu}(z) t_{a}
  \right]
\end{equation}
enter the quark-antiquark matrix elements, which accumulate
non-perturbative information on the distribution of partons
inside a hadron participating in high-energy collisions.
In these parton distribution functions Wilson lines arise due to the
resummation of gluon exchanges between the hard and the soft
part of the (factorized) process, while the integration contour
$\Gamma$ is defined by the hard subprocess.
One the other hand, from the field-theoretical point of view, Wilson
lines restore the gauge invariance of nonlocal two-fermion operators.

It is known that the renormalization of the contour-dependent operators
with Wilson-line obstructions (cusps, or self-intersections) cannot be
performed by the ordinary $R-$operation alone, but requires an
additional renormalization factor depending on the cusp angle
\cite{Pol80,CD80,Aoy81,KR87}:
\begin{equation}
Z (\chi) = \[\left\<0 \left| \pa \exp\[ig \int_{\Gamma_\chi} d\z^\m
   \ \hat A^a_\m (\z)\] \right|0\right\> \]^{-1} \ .
\end{equation}
From this expression, one can find the corresponding
(cusp-angle-dependent) anomalous dimension:
\begin{equation}
  \gamma_{\rm cusp}
=
  \halb \ \frac{1}{Z (\chi)} \ \m \frac{\pd \a_s (\m)}{\pd\m}
  \ \frac{\pd Z(\chi) (\m, \a_s (\m); \e)} {\pd\a_s} \ .
\end{equation}
The ultraviolet (UV) anomalous dimensions of TMD PDFs are on the
focus of the present report, since they accumulate the main
characteristics of Wilson lines in \textit{local} form, while the
gauge contours themselves are \textit{global} objects and, therefore,
complicated to be treated within a local-field theory framework.

Parton distribution functions play a principal role in QCD
phenomenology \cite{Col03,BR05,Col08}.
In inclusive processes, such as DIS, the standard (integrated) PDFs
are used, which depend on the longitudinal fraction of the momentum,
$x$, and on the scale of the hard subprocess $Q^2$.
The completely gauge invariant (with the Wilson line inserted)
definition of integrated PDFs reads \cite{CS81}
\begin{equation}
  f_{i}(x)
=
  \frac{1}{2} \int \frac{d\xi^-}{2\pi}\
  \ex^{- i k^+ \xi^-} { \langle h(P) | } \bar \psi_i
  (\xi^-, \vecc 0_\perp) [\xi^-, 0^-]\gamma^+
  \psi_i(0^-,\vecc 0_\perp) { |h(P)\rangle }
\end{equation}
and its renormalization properties are described by the DGLAP
evolution equation
\begin{equation}
  \mu \frac{d}{d\mu} f_{i} (x, \mu)
=
  \sum_j \int_x^1\! \frac{dz}{z} \ P_{ij}
  \left(\frac{x}{z}\right) f_{j} (z, \mu) \ ,
  \label{eq:dglap}
\end{equation}
where $P_{ij}$ is the DGLAP integral kernel.

The study of semi-inclusive processes, such as SIDIS, or the Drell-Yan
process, where the transverse momentum of the produced hadrons can be
observed, requires the introduction of more complicated quantities,
so-called unintegrated, or transverse-momentum dependent, PDFs.
In this case, the integration over the transverse component of the
parton's momentum $\vecc k_\perp$ is not performed.
Their gauge-invariant definition looks like
\cite{JY02,JMY04,BJY03,BMP03}
($\!\xi^{+}=0$)
$$
   f_{q/q}(x, \mbox{\boldmath$k_\perp$})
 =
  \frac{1}{2}
   \int \frac{d\xi^- d^2
   \vecc \xi_\perp}{2\pi (2\pi)^2}
   {\rm e}^{- i k^+ \xi^- +i \mbox{\footnotesize \boldmath$k_\perp$}
   \cdot \mbox{\footnotesize \boldmath$\xi_\perp$}}
   \left\langle  q(p) |\bar \psi (\xi^-, \xi_\perp)
   [\xi^-, \mbox{\boldmath$\xi_\perp$};
   \infty^-, \mbox{\boldmath$\xi_\perp$}]^\dagger \right.
$$
\begin{equation}
\left.
\times
   [\infty^-, \mbox{\boldmath$\xi_\perp$};
   \infty^-, \mbox{\boldmath$\infty_\perp$}]^\dagger \gamma^+
[\infty^-, \mbox{\boldmath$\infty_\perp$};
   \infty^-, \mbox{\boldmath$0_\perp$}]
   [\infty^-, \mbox{\boldmath$0_\perp$}; 0^-,\mbox{\boldmath$0_\perp$}]
   \psi (0^-,\mbox{\boldmath$0_\perp$}) |q(p)\right\rangle \
   \ .
\label{eq:tmd_definition}
\end{equation}
Formally, the integration over the transverse component of the parton's
momentum is expected to yield the integrated distribution
\begin{equation}
  \int\! d^2 k_\perp f_i (x, \vecc k_\perp)
=
  f_{i/h} (x)\ .
\end{equation}

However, the above definition cannot be considered as a final one and
in fact it has to be modified.
The reason is that in this case, extra (rapidity) divergences
arise---which are associated with the known features of the light-cone
gauge or the light-like Wilson lines---that cannot be removed by
ordinary UV renormalization alone \cite{CS81,Col08,CRS07,Bacch08,CS07}.
Note that in the integrated case these divergences, though they do
appear at the intermediate steps of the calculation, they are absent
in the final result due to the mutual cancelation between real and
virtual gluon contributions.
A further complication is that the reduction to the integrated case
cannot be performed straightforwardly: the formal integration does
not reproduce the correct result (i.e., the DGLAP kernel) because of
additional uncanceled UV divergences.

In this presentation, we report on a generalized renormalization
procedure \cite{CS07} for non-lightlike Wilson lines (which is akin
to the subtractive method of Ref.\ \cite{CH00,Hau07}---see also
\cite{CM04}) in order to remove the extra divergences by an
additional ``soft'' factor, defined by the vacuum average of particular
Wilson lines.
This allows us to perform the necessary modifications of TMD PDFs in
the most economic way.
To this end, we calculate the anomalous dimension of the TMD PDF
(in fact, we calculate the distribution of a ``quark in a quark'') in
the light-cone gauge and identify the extra UV divergences which
generate an additional anomalous dimension.
Then, we  perform a generalized renormalization procedure of the TMD
PDF, similar to the renormalization of contour-dependent operators with
cusped or self-intersecting gauge contours \cite{Pol80, CD80}.
This renormalization cancels all undesirable divergences and yields
a completely gauge invariant definition of TMD PDFs.

\section{Analysis of the leading-order UV divergences}
\label{sec:UV-div}

\begin{figure}
\centering
\vspace*{0mm} \scalebox{1.3}{\includegraphics[width=0.15\textwidth,angle=90]{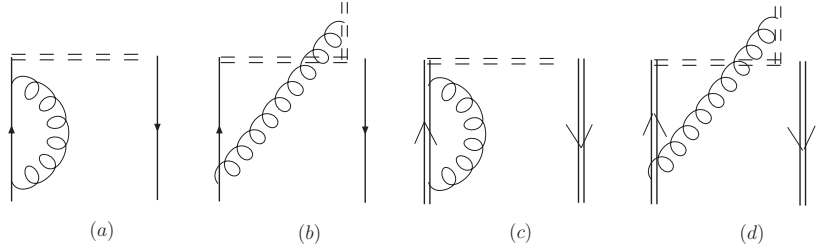}}
 \caption{One-gluon exchanges in the TMD PDF in the light-cone gauge.
  Diagrams $(a)$ and $(b)$ give rise to UV divergences, while
  $(c)$ and $(d)$ correspond to the soft factor,
  cf.\ (\ref{eq:soft_factor_1}).
  Double lines denote gauge links and curly lines gluon propagators.
  The Hermitian conjugate, i.e., ``mirror'', diagrams are omitted.
}
\label{fig:1}
\end{figure}

The one-gluon exchanges, contributing to the UV-divergences,
are described by the diagrams $(a)$ and $(b)$ in Fig.\ \ref{fig:1}.
In the light-cone gauge, extra rapidity divergences arise
owing to the $q^+$-pole in the gluon propagator:
\begin{equation}
  D^{\m\n}_{\rm LC} (q)
= \frac{-i}{q^2}
  \[g^{\m\n} - {\frac{q^\m n^{-\n}}{[q^+]}
             - \frac{q^\n n^{-\m} }{[q^+]}}
  \] \ ,
\end{equation}
where $[q^+]$ stands for an undefined denominator.
We consider the following pole prescriptions:
\begin{equation}
  \frac{1}{[q^+]}_{\rm PV}
=
  \halb \( \frac{1}{q^+ + i \eta} + \frac{1}{q^+ - i \eta} \) \ \
\hbox{and} \ \
  \frac{1}{[q^+]}_{\rm Adv/Ret}
=
  \frac{1}{q^+ \mp i \eta}  \ ,
  \label{eq:pole}
\end{equation}
where $\eta$ is small but finite.
To control UV singularities, dimensional regularization is used.
Another possible prescription, namely, the Mandelstam-Leibbrand
(ML) one, will be considered in future work.

The UV divergent part of the diagrams $(a)$ and $(b)$, depicted in
Fig.\ \ref{fig:1} (without their ``mirror'' contributions), is
\begin{equation}
  {\S}^{UV}_{\rm left} (p, \a_s ; \e)
=
  {
   - \frac{\a_s}{\pi}C_{\rm F} \ \frac{1}{\e}
   \[- \frac{3}{4} - \ln \frac{\eta}{p^+} + \frac{i\pi}{2}
    + i \pi \ C_\infty \] + \a_s C_{\rm F} \ \ \frac{1}{\e}
    \[i C_\infty\]
  } \ ,
\label{eq:left}
\end{equation}
where
$C_{\rm F}=\left(N_{\rm c}^{2}-1\right)/2N_{\rm c}=4/3$
and the numerical factor $C_{\infty}$ accumulates the
pole-prescription uncertainty, defined by
\begin{equation}
 C_\infty^{\rm Adv} = 0 \ , \ C_\infty^{\rm Ret} = -1 \ ,
 C_\infty^{\rm PV} = -1/2 \ .
\end{equation}
One appreciates that the contribution of the transverse gauge
link at light-cone infinity---diagram 1($b$)---suffices to
cancel the dependence on the pole prescription.

Turn now to the important issue of {\it time-reversal-odd}
effects that appear already in Eq.\ (\ref{eq:left}) and are
responsible for single-spin asymmetries \cite{Mul08}.
It is expected that $T$-odd effects arise when the dependence on
the intrinsic transverse motion of partons is taken into account,
e.g., in semi-inclusive processes, like SIDIS (or DY).
In covariant gauges, this effect originates from the Wilson lines in
the operator definition of the TMD PDFs.
On the other hand, in the axial (light-cone) gauge, our analysis
demonstrates that $T-$odd phenomena reveal themselves via the
direction to go around the pole.
In fact, the imaginary term,
\begin{equation}
 {\rm Im} \ {\S}^{UV}_{\rm left}
=
 - \frac{\a_s}{2\e}C_{\rm F} \
\end{equation}
in Eq.\ (\ref{eq:left}) stems from the infinitesimal deformation
of the integration contour to circumvent the pole in the gluon
propagator subject to the pole prescriptions in (\ref{eq:pole}).
It corresponds to the imaginary term one would obtain with lightlike
Wilson lines in a covariant gauge.
In this latter case, the leading term in the Wilson line produces,
after Fourier transforming it, has a similar $q^+$-pole in the
denominator:
\begin{equation}
  \int_0^\infty\! d\xi^- A^+ (\xi^-, 0^+, \vecc 0_\perp)
=
  \int\! d^4 q \tilde A^+ (q)
  \int_0^\infty\! d\xi^- \ex^{-i (q^+ - i\eta)\xi^-}
=
  \int\! d^4 q \tilde A^+ (q) \frac{-i}{q^+ - i\eta} \ .
   \label{eq:cov}
\end{equation}
Taking into account that $T-$reversal corresponds to the inversion
of the Wilson line's direction by flipping the sign in the denominator
from $\eta \to - \eta$, one may conclude that the $T-$odd effects
in the TMD PDFs are already ingrained in their {\it local} RG
properties and are revealed by their anomalous dimensions.
Note that the imaginary terms in the anomalous dimensions can be
attributed to the contributions of gluons in the Glauber regime, where
their momenta are mostly transverse \cite{KR87}.
Indeed, it was recently shown (in the Soft Collinear Effective Theory)
that exactly the Glauber gluons contribute most to the transverse gauge
link underlying $T-$odd effects in the light-cone gauge \cite{IdMa08}.

Taking into account the ``mirror'' contributions (termed ``right'' in
the equation to follow), one gets the total real UV divergent part:
\begin{equation}
  \S_{\rm tot}^{UV} (p, \a_s (\m) ; \e )
=
  \S_{\rm left} +  \S_{\rm right} =
  {  - \frac{ \a_s}{4\pi}C_{\rm F} \   \frac{2}{\e}
  \(- 3 - 4 \ln \frac{\eta}{p^+} \) } \ .
\label{eq:tot_uv}
\end{equation}

Thus, the one-loop anomalous dimension reads
\begin{equation}
  \gamma_{\rm LC}
=
  \gamma_{\rm smooth} - {  \d \gamma } \ \ , \ \ \gamma_{\rm smooth}
=
  {  \frac{3}{4} \frac{\a_s}{\pi}C_{\rm F} } + O(\a_s^2) \ .
\end{equation}
The defect of the anomalous dimension
\begin{equation}
  \d \gamma
=
  - \frac{ \a_s}{\pi}C_{\rm F} \  \ln \frac{\eta}{p^+} \ ,
\end{equation}
marks the deviation of the calculated quantity from the
anomalous dimension of the two-quark operator with the smooth
(i.e., direct) gauge connector.
The connector-corrected fermion propagator \cite{CD80,Ste83} has an
anomalous dimension that is twice the anomalous dimension of the
fermion field operator.
On the other hand, $\gamma_{\rm LC}$ contains an undesirable
$p^+$-dependent term that should be removed by an appropriate
procedure.
Note that $p^+ = (p \cdot n^-) \sim \cosh \chi$ defines, in fact,
an angle $\chi$ between the direction of the quark momentum
$p_\mu$ and the light-like vector $n^-$.
In the large $\chi$ limit, $\ln p^+ \to \chi , \ \chi \to \infty$.
Thus, we conclude that the defect of the anomalous dimension,
$\delta \gamma$, can be identified with the one-loop cusp anomalous
dimension \cite{KR87}.

\section{Modified definition of the TMD PDF}
\label{sec:mod-TMD-PDF}

Using the above observation as a hint and taking into account the
renormalization properties of Wilson lines with obstructions,
discussed in the Introduction, we now compute the extra
renormalization constant associated with the soft counter term
\cite{CH00} and show that it can be expressed in terms of a vacuum
expectation value of a specific gauge link.
Hence, in order to cancel the anomalous dimension defect
$\delta \gamma$, we introduce the counter term
\begin{equation}
  R
\equiv
 \Phi (p^+, n^- | 0) \Phi^\dagger (p^+, n^- | \xi) \ ,
\label{eq:soft_factor_1}
\end{equation}
where
\begin{equation}
  \Phi (p^+, n^- | \xi )
 =
  \left\langle 0
  \left| {\cal P} \exp\Big[ig \int_{\Gamma_{\rm cusp}}d\zeta^\mu
  \ t^a A^a_\mu (\xi + \zeta)\Big]
  \right|0
  \right\rangle
\label{eq:soft_definition}
\end{equation}
and evaluate it along the non-smooth, off-the-light-cone integration
contour
\begin{equation}
\Gamma_{\rm cusp} : \ \ \zeta_\mu
=
  \{ [p_\mu^{+}s \ , \ - \infty < s < 0] \
 \cup \ [n_\mu^-  s' \ ,
  \ 0 < s' < \infty] \ \cup \
  [ \mbox{\boldmath$l_\perp$} \tau , \, \ 0 < \tau < \infty ] \}
\label{eq:gpm}
\end{equation}
with $n_\mu^-$ being the minus light-cone vector.

The one-loop gluon virtual corrections, contributing to the UV
divergences of the soft factor $R$, are shown in Fig.\ \ref{fig:1}
(diagrams $(c)$ and $(d)$).
For the UV divergent term we obtain
\begin{equation}
  \Sigma_{R}^{\rm UV}
=
  - \frac{ \alpha_s}{\pi} C_{\rm F} \ 2 \left(  \frac{1}{\epsilon} \
  \ln \frac{\eta}{p^+} - \gamma_E + \ln 4 \pi \right)
\end{equation}
and observe that this expression is equal, but with opposite sign,
to the unwanted term in the UV singularity related to the cusped
contour calculated above.

Therefore, we propose to  redefine the conventional TMD PDF and
absorb the soft counter term in its definition:
$$
   f_{q/q}^{\rm mod}(x, \mbox{\boldmath$k_\perp$})
\! = \!
  \frac{1}{2} \!
   \int \! \frac{d\xi^- d^2 \!
   \vecc \xi_\perp}{2\pi (2\pi)^2}
   {\rm e}^{- i k^+ \xi^- + i
   {\vecc k_\perp} \cdot {\vecc \xi_\perp}}
   \bigl\langle  q(p) |\bar \psi (\xi^-, \xi_\perp)
   [\xi^-, \mbox{\boldmath$\xi_\perp$};
   \infty^-, \mbox{\boldmath$\xi_\perp$}]^\dagger
   [\infty^-, \mbox{\boldmath$\xi_\perp$};
   \infty^-, \mbox{\boldmath$\infty_\perp$}]^\dagger
$$
\begin{equation}
   \times \gamma^+[\infty^-, \mbox{\boldmath$\infty_\perp$};
   \infty^-, \mbox{\boldmath$0_\perp$}]
   [\infty^-, \mbox{\boldmath$0_\perp$}; 0^-,\mbox{\boldmath$0_\perp$}]
   \psi (0^-,\mbox{\boldmath$0_\perp$}) |q(p)\bigr\rangle
   \cdot
   R (p^+, n^-) \ .
\label{eq:tmd_re-definition}
\end{equation}

One immediately verifies that the integration over the transverse
momentum $\vecc k_\perp$ yields the integrated PDF
\begin{equation}
   \int\! d^{\omega-2} \vecc k_\perp
   f_{i/a}^{\rm mod} (x, \vecc k_\perp ; \mu , \eta)
   =
   f_{i/a} (x, \mu) \ ,
\end{equation}
which obeys the DGLAP equation (\ref{eq:dglap}).

\section{Conclusions}
\label{sec: concl}

To conclude, we found \cite{CS07} that the additional UV divergences
in the TMD PDFs are related to the renormalization effect on the
junction point of Wilson lines, when they contain transverse segments
extending to light-cone infinity.
The anomalous dimension ensuing from these divergences coincides
with the (one-loop) universal cusp anomalous dimension
\cite{KR87}.
A modified definition of the TMD PDFs was proposed, which
contains a soft counter term in the sense of Collins and Hautmann
\cite{CH00} which is a path-ordered exponential factor evaluated along
a particular gauge contour with a cusp.
The anomalous dimension associated with the renormalization of this
nonlocal operator compensates the anomalous-dimension artifact and
ensures that integrating over the parton transverse momentum, one finds
a PDF satisfying the DGLAP evolution equation.
Moreover, the anomalous dimension of the modified TMD PDF respects the
Slavnov-Taylor identities and resembles the one-loop expression one
finds for a TMD PDF with a connector insertion \cite{Ste83} (see also
\cite{CD80}), i.e., the direct Wilson line between the two quark
fields.
The cusp-like junction point is ``concealed'' by light-cone infinity,
and reveals itself only \textit{after} renormalization as a phase
entanglement \cite{CS07} akin to the ``intrinsic'' Coulomb phase,
found before in QED \cite{JS90}, and being codified in the (one-loop)
cusp anomalous dimension.
The implications of a more accurate definition of TMD PDFs are far
reaching, ranging from more precise analyses of various experimental
data on hard-scattering cross sections to the development of more
accurate Monte Carlo event generators.

\section{Acknowledgements}

This work was supported in part by the Alexander von Humboldt-Stiftung,
the Deutsche Forschungsgemeinschaft
under grant 436 RUS 113/881/0,
the RF Scientific Schools grant 195.2008.9,
the Heisenberg--Landau Programme 2008, and the INFN.


\begin{thebibliography}{99}

\bibitem{Pol80}
  A.M.~Polyakov,
  Nucl.\ Phys.\ B {\bf 164}, 171 (1980).

\bibitem{CD80}
  N.S.~Craigie and H.~Dorn,
  Nucl.\ Phys.\ B {\bf 185}, 204 (1981).

\bibitem{Aoy81}
  S.~Aoyama,
  Nucl.\ Phys.\ B {\bf 194}, 513 (1982).

\bibitem{KR87}
  G.P.~Korchemsky and A.V.~Radyushkin,
  Nucl.\ Phys.\ B {\bf 283}, 342 (1987).

\bibitem{Col03}
   J.C.~Collins,
   Acta Phys.\ Pol.\ B {\bf 34}, 3103 (2003).

\bibitem{BR05}
  A.V.~Belitsky and A.V.~Radyushkin,
  Phys.\ Rept.\ {\bf 418}, 1 (2005).

\bibitem{Col08}
  J.~Collins,
  arXiv:0808.2665 [hep-ph].

\bibitem{CS81}
   J.C.~Collins and D.E.~Soper,
   Nucl.\ Phys.\ {\bf B193}, 381 (1981);
   {\bf B213}, 545 (E) (1983).

\bibitem{JY02}
  X.~Ji and F.~Yuan,
  Phys.\ Lett.\  B {\bf 543}, 66 (2002).

\bibitem{JMY04}
  X.~Ji, J.~Ma, and F.~Yuan,
  Phys.\ Rev.\ D {\bf 71}, 034005 (2005).

\bibitem{BJY03}
  A.V.~Belitsky, X.~Ji, and F.~Yuan,
  Nucl.\ Phys.\ B {\bf 656}, 165 (2003).

\bibitem{BMP03}
   D.~Boer, P.J.~Mulders, and  F.~Pijlman,
   Nucl.\ Phys.\ {\bf B667}, 201 (2003).

\bibitem{CRS07}
  J.C.~Collins, T.C.~Rogers, and A.M.~Stasto,
  Phys.\ Rev.\ D {\bf 77}, 085009 (2008).

\bibitem{Bacch08}
  A.~Bacchetta, D.~Boer, M.~Diehl and P.~J.~Mulders,
  JHEP {\bf 0808}, 023 (2008).

\bibitem{CS07}
  I.O.~Cherednikov and N.G.~Stefanis,
  Phys.\ Rev.\ D {\bf 77}, 094001 (2008);
  Nucl.\ Phys.\ B {\bf 802}, 146 (2008).

\bibitem{CH00}
   J.C.~Collins and F.~Hautmann,
   Phys.\ Lett.\ B {\bf 472}, 129 (2000).

\bibitem{Hau07}
  F.~Hautmann,
  Phys.\ Lett.\ B {\bf 655}, 26 (2007).

\bibitem{CM04}
  J.C.~Collins and A.~Metz,
  Phys.\ Rev.\ Lett.\ {\bf 93}, 252001 (2004).

\bibitem{Mul08}
  P.J.~Mulders,
  arXiv:0810.3772 [hep-ph].

\bibitem{IdMa08}
  A.~Idilbi and A.~Majumder,
  arXiv:0808.1087 [hep-ph].

\bibitem{Ste83}
  N.G.~Stefanis,
  Nuovo Cim.\ A {\bf 83}, 205 (1984).

\bibitem{JS90}
  R.~Jakob and N.G.~Stefanis,
  Annals Phys.\ {\bf 210}, 112 (1991).

\end{thebibliography}
\end{document}